\documentclass[letter]{aa}
\usepackage{txfonts,astro,natbib,graphicx,amssymb,color}
\usepackage[normalem]{ulem}

\begin{document}
\def\sigt{\ensuremath{\sigma_T}}
\def\sigtdv{\ensuremath{\sigma_W}}
\def\tdust{\ensuremath{T_{\rm dust}}}
\def\ra{\ensuremath{\rightarrow}}
\def\pdf{\textit{pdf}}
\def\phb#1{\textbf{PHB:~{#1}}}
\def\corr#1#2{\red{\sout{#1}}{\,\textbf{#2}}}
\def\pv{\ensuremath{p-v}}
\def\answer#1{\textbf{#1}}

\title{Intermittency of interstellar turbulence: Parsec-scale coherent
  structure of intense velocity-shear\thanks{Based on observations
    carried out with the IRAM-30~m telescope. IRAM is supported by
    INSU-CNRS/MPG/IGN.}}

\author{%
  P.~Hily-Blant\inst{1}
  \and
  E.~Falgarone\inst{2}
}

\offprints{P. Hily-Blant, \email{pierre.hilyblant@obs.ujf-grenoble.fr}}

\institute{LAOG, CNRS \& Universit\'e Joseph Fourier, UMR 5571, 414
  Rue de la Piscine BP 53 F-38041 Grenoble Cedex 09 \and LRA/LERMA,
  CNRS \& \'Ecole normale sup\'erieure \& Observatoire de Paris, UMR
  8112, 24 rue Lhomond, 75231 Paris Cedex 05, France }

\date{Received / Accepted}

\abstract{} {Guided by the duality of turbulence (random versus
  coherent motions), we seek coherent structures in the turbulent
  velocity field of molecular clouds, anticipating their importance in
  cloud evolution.}  {We analyse a large map (40' by 20') obtained
  with the HERA multibeam receiver (IRAM-30m telescope) in a high
  latitude cloud of the Polaris Flare at an unprecedented spatial
  (11\arcsec) and spectral (0.05\kms) resolutions in the
  \twCO\jtwo\ line.}  {We find that two parsec-scale components of
  velocities differing by $\sim 2$~\kms, share a narrow interface
  ($<0.15$~pc) that appears as an elongated structure of intense
  velocity-shear, $\sim 15$ to 30~\kmspc.  The locus of the extrema of
  line--centroid-velocity increments (E-CVI) in that field follows
  this intense-shear structure as well as that of the
  \twCO\jtwo\ high-velocity line wings.  The tiny spatial overlap in
  projection of the two parsec-scale components implies that they are
  sheets of CO emission and that discontinuities in the gas properties
  (CO enrichment and/or increase of gas density) occur at the position
  of the intense velocity shear.}  {These results disclose spatial and
  kinematic coherence between scales as small as 0.03~pc and parsec
  scales. They confirm that the departure from Gaussianity of the
  probability density functions of E-CVIs is a powerful statistical
  tracer of the intermittency of turbulence. They disclose a link
  between large scale turbulence, its intermittent dissipation rate
  and low-mass dense core formation.}


\authorrunning{Hily-Blant \& Falgarone}

\titlerunning{Coherent structure of intense velocity-shear}
\maketitle

\section{Introduction}

Because it is supersonic, magnetized and develops in a multiphase
medium, interstellar turbulence is expected to differ from turbulence
in laboratory flow experiments or in state-of-the-art numerical
simulations, \eg\ \cite{chanal2000} for recent experiments in gaseous
helium and \cite{mininni2006a} or \cite{alexakis2007} for MHD
simulations.  Nonetheless, it may carry some universal properties of
turbulence, such as space-time intermittency \citep[for a review
  see][]{anselmet2001}.  Of particular interest to star formation, is
the behavior of turbulence dissipation.  In a series of papers, we
have shown that the \twCO\jone\ line--centroid-velocity increments
(CVI) in translucent molecular gas have non-Gaussian statistics more
pronounced at small scale. The extreme CVI (E-CVI) responsible for the
non-Gaussian tails of their probability density functions (\pdf) form
elongated coherent structures over 0.8~pc.  These structures have been
tentatively identified with regions of intense
velocity-shear~\footnote{In the following, ``shear'' is used instead
  of gradient to \answer{emphasize} that the observations provide
  cross-derivatives of the line--centroid-velocities (CV) (\ie\ the
  displacement, in the plane-of-the-sky (POS), is perpendicular to the
  projection axis of the velocities).}
%
\answer{and enhanced local dissipation rate, based on} their thermal,
dynamical, and chemical properties. These pure velocity-structures do
not follow those of dense gas, they tend to be parallel to the
magnetic field orientation, they are associated with gas warmer
($\tkin>25$~K) than the bulk of the gas \citep[][ hereafter Paper~I
  and Paper~II]{hilyblant2007ii,hilyblant2008cvi}, and they bear
chemical signatures ascribed to a warm chemistry not driven by UV
photons \citep{falgarone2006hcop, godard2009}.  Last, in one such
E-CVI structure, Plateau de Bure Interferometer (PdBI) observations
disclose several sub-structures of intense velocity-shear at scales as
small as 6~\answer{milli-parsec (mpc)} \citep[][ hereafter
  FPH09]{falgarone2009pdbi}. This suggests that turbulent molecular
clouds harbour coherent velocity-shear structures from 6~mpc to
800~mpc.

We have increased the dynamic range by a factor 8 compared with
Paper~II, by mapping four times larger an area in the Polaris Flare
with twice better a spatial resolution. The aim is to further explore
the range of scales over which the spatial coherence of these
intense-shear structures is found.  These are the first large scale
observations performed at high-angular resolution and high spectral
resolution in translucent gas. The observations and the results are
described in Section 2 and 3.  We briefly discuss possible
interpretations and the nature of these structures in light of recent
numerical simulations (Section 4).

\section{Observations}
\begin{figure*}[t]
  \centering
  \includegraphics[height=0.4\hsize,angle=-90]{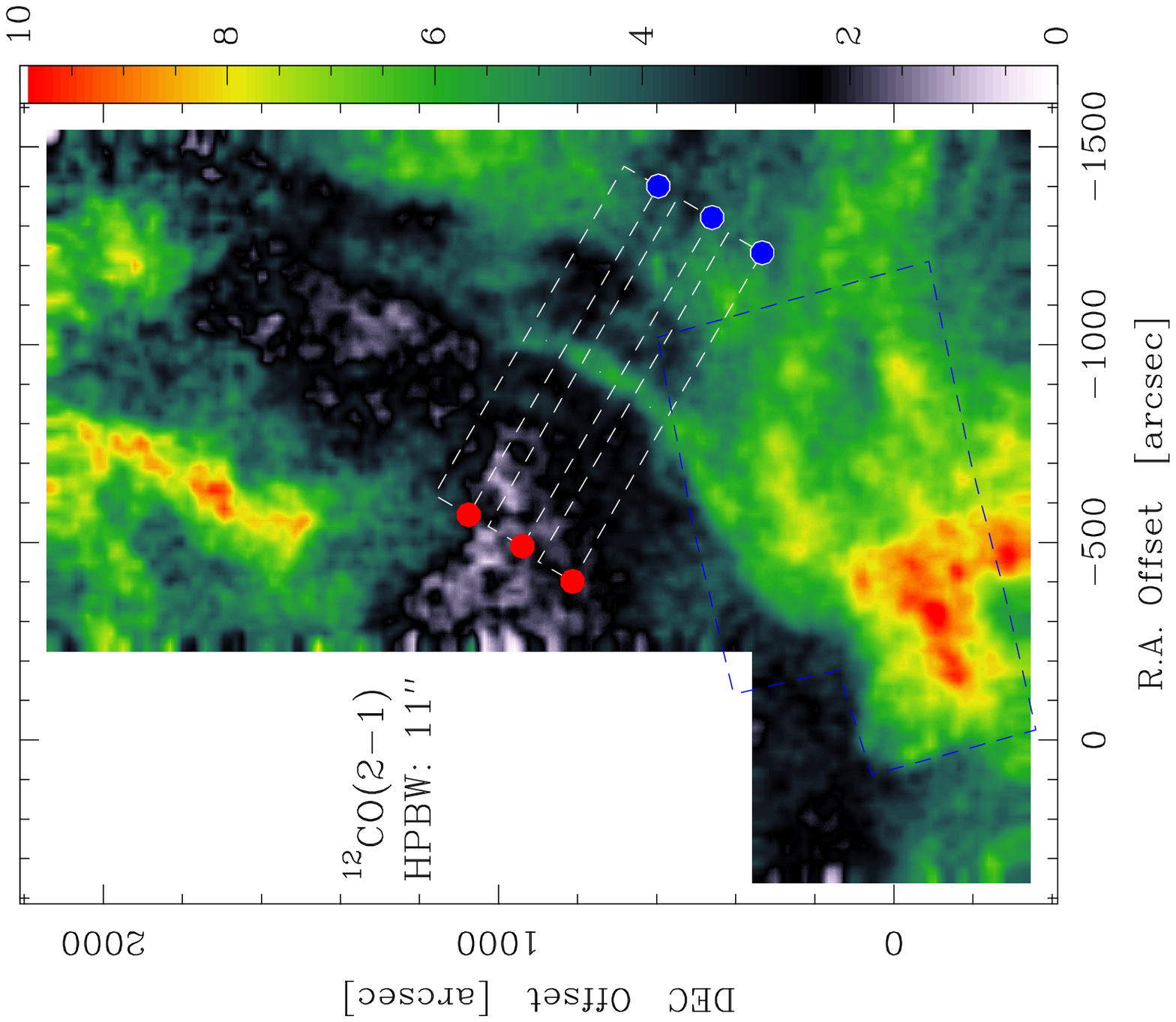}\hskip0.05\hsize
  \includegraphics[width=0.45\hsize,angle=-90]{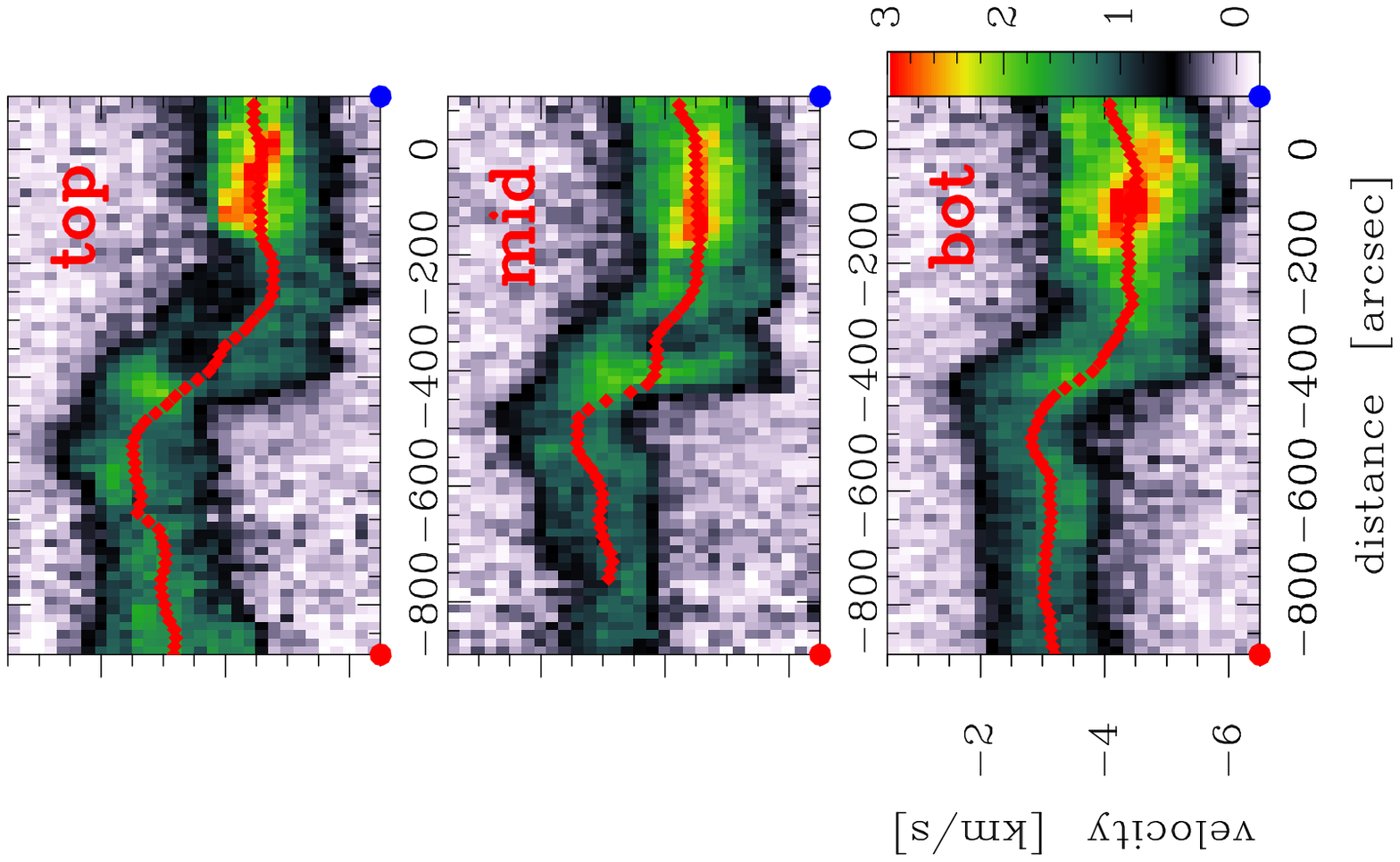}
  \caption{\textit{Left:} Integrated intensity map (\kkms, \tant
    scale) smoothed to 15\arcsec. The white dashed boxes show the
    areas used to build the average \pv\ cuts (right panels). The blue
    rectangle (dashed line) delineates the previous field of
    Paper~I. \textit{Right:} \pv\ cuts in the 3 boxes shown in left
    panel (\tant\ scale, distance measured in \arcsec\ with arbitrary
    origin). The line CV are shown in red. }
  \label{fig:tdv}
\end{figure*}

Observations of the \twCO\jtwo\ line were carried out at the IRAM-30m
telescope with the 1.3mm multibeam heterodyne receiver HERA
\citep{schuster2004} during August 2007 and January 2008, under good
weather conditions. The map covers 0.3~deg$^2$ and consists of 9
submaps of $10\arcmin\times10\arcmin$, each of which was observed in
two orthogonal scanning directions to minimize striping due to gain or
atmosphere variations. The final map encompasses the fields
successively observed by \cite{falgarone1998kp} and in Paper~I shown
as a dashed box in Fig.\ref{fig:tdv} and is $\sim 2\times 1$~pc at the
adopted distance of the source (150~pc).  Data were acquired in the
powerful on-the-fly (OTF) frequency-switched mode ($4\arcsec\, s^{-1}$
scanning velocity, $1s$ time sampling, $4\arcsec$ spatial sampling in
both directions, 13.8~MHz frequency throw), using the VESPA
autocorrelator facility as backends. A total of 1.5\tdix{6}
\answer{raw} spectra was recorded in 80~hours of telescope time with a
spectral resolution of 0.05~\kms. Data were reduced with the new
\texttt{CLASS90} software optimized for OTF
\citep{IRAM_report_2005-1}. The instrumental response was canceled by
subtracting linear baselines to each original spectrum, which were
then convolved by a Gaussian kernel ($1/3\,HPBW$) and gridded on a
regular grid with $0.5 HPBW$ sampling. The final data cube was then
smoothed to 15\arcsec\ and 0.1~\kms\ resolutions to improve the
signal-to-noise ratio. The typical rms in each final pixel is
$1\sigma=0.5$~K in 0.1~\kms\ channels.

\section{Results}

\subsection{Space and velocity maps}

\begin{figure}
  \centering
  \includegraphics[width=0.7\hsize,angle=0]{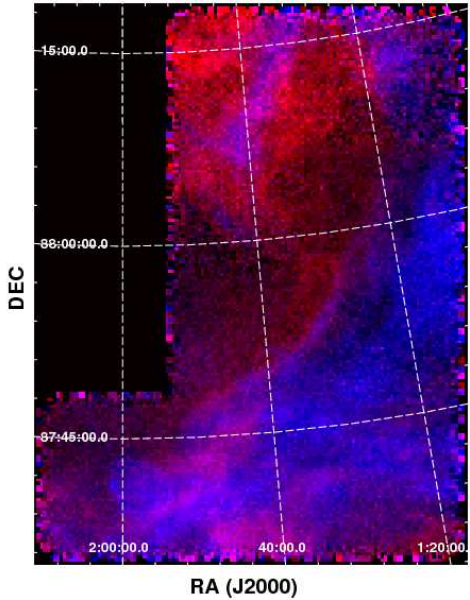}
  \caption{\twCO\jtwo\ integrated intensity in two adjacent velocity
    ranges: [-6.5:-3.5]~\kms\ in blue and [-3.5:-0.5]~\kms\ in red.
    At a distance of 150~pc, 20' correspond to 0.9~pc.}
  \label{fig:rgb}
\end{figure}

The \twCO\jtwo\ integrated emission is displayed in Fig.~\ref{fig:tdv}
(left panel) with three position-velocity (\pv) diagrams (right
panels) made along the NE-SW boxes shown.  A sharp variation of
velocity, from $\sim-3$~\kms\ to $\sim-5$~\kms\ (from NE to SW) occurs
over a layer thinner than a few 100\arcsec\ in
projection. Fig.\ref{fig:rgb} that displays the integrated emission in
two adjacent velocity intervals, at high (HV) [-3.5, -0.5]~\kms\ and
low velocity (LV) [-6.5, -3.5]~\kms, stresses a remarkable
characteristic of that field: the edges of the LV and HV components
follow each other closely in projection over more than $\sim1$~pc.  It
is then most unlikely that they be unrelated pieces of gas along the
line of sight: they have to be in contact.

The second remarkable characteristic of that field is the following:
while the emissions in the LV and HV components are extended, their
spatial overlap (the pink areas of Fig.\ref{fig:rgb}, also exemplified
in the \pv\ diagrams) is limited to narrow filamentary regions in
projection.  It is most visible between $\delta=87\deg45'$ and
$88\deg$ (thus over $\sim 1$~pc) where it does not split into several
substructures.  Now, if the LV and HV components were parsec-scale
volumes, their interface would appear thin over $\sim 1$~pc only if
viewed edge-on (within $\pm 5$\deg\ for a projected size less than one
tenth of its real size), a case that we rule out on statistical
grounds. We therefore infer that the \twCO\jtwo\ HV and LV components
are {\it layers} rather than {\it volumes} and that their interface is
1-dimensional rather than 2-dimensional. This ensures that under any
viewing angle the two extended velocity components present a narrow
interface.

The slope of the variations of the line CV drawn on the \pv\ diagrams
provides a measurement of the velocity shear between the two
components.  On each cut shown, there is an average shear of $\approx
13$~\kmspc\ (a velocity variation of 2~\kms\ over $\approx
0.15$~\pc). Steeper slopes are also visible and locally provide higher
shears up to 30~\kmspc\ (1~\kms\ over 0.03~\pc) in the middle cut.
These values are more than one order of magnitude larger (within the
uncertainties due to projections) than the average value of
1~\kmspc\ estimated at the parsec scale in molecular clouds
\citep{goldsmith1985}. The velocity field therefore significantly
departs, at small scales, from predictions based on the generally
adopted scaling laws in molecular clouds. If velocity fluctuations
over a scale $l$ increase as $\delta v_l \propto l^{1/2}$,
velocity-shears should increase by no more than $33^{1/2}\approx 5.7$
between 1~\pc\ and 0.03~\pc.

A closer inspection of the \pv\ diagrams shows that \textit{(i)} the
sharpest variations of the line CV occur between two line-wings
appearance (above $-2.0$~\kms\ for the HV wing and below
$-5.5$~\kms\ for the LV wing), \textit{(ii)} the separation between
the LV and HV wings steepens from top (0.1~pc) to bottom (0.03~pc),
and \textit{(iii)} the layer of largest velocity shear coincides with
the lane of enhanced \twCO\jtwo\ emission visible in
Fig.~\ref{fig:tdv} at the center of each cut.
  
\subsection{Distribution of E-CVI}

\begin{figure*}
  \centering
  \includegraphics[width=\hsize]{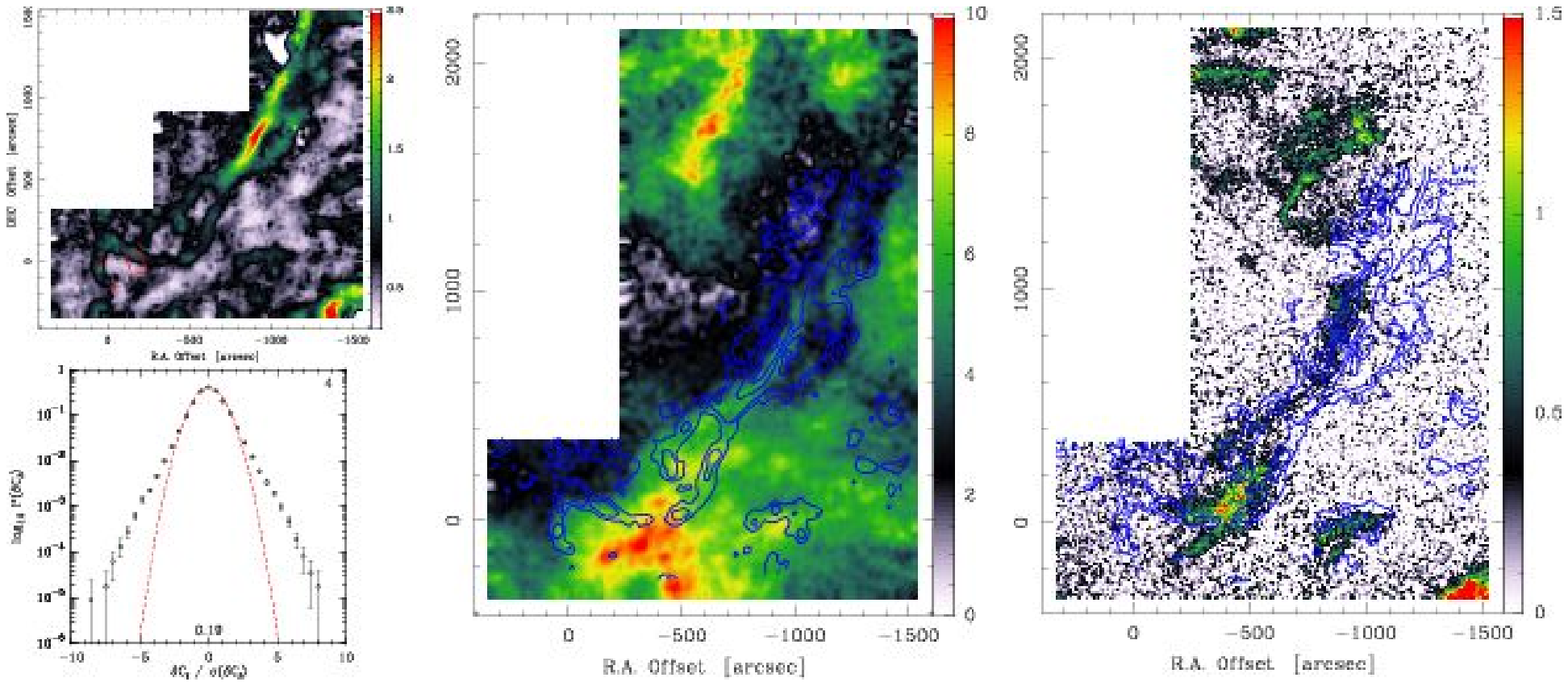}
  \caption{\textit{Left:} \answer{Map of the CVI (top panel, color
      scale in \kms) computed for a lag of 4 pixels or 60\arcsec, and
      normalized \pdf\ (circles, bottom panel) compared with a
      Gaussian distribution ($\sigma=0.19$~\kms, red)}. The red
    crosses (top panel) indicate the position of the dense cores from
    \cite{heithausen2002}. The rectangle delineates the PdBI field of
    \cite{falgarone2009pdbi}. \textit{Middle:} E-CVI (blue contours)
    overplotted on the integrated intensity of
    Fig.~\ref{fig:tdv}. \textit{Right:} E-CVI (blue contours)
    overplotted on the intensity integrated in the red wing interval
    [-2:-0.5]~\kms\ \citep{hilyblant2007ii}.}
  \label{fig:cvi}
\end{figure*}

Following \cite{lis1996,pety2003} and Paper~II, we have built the
\pdf\ \answer{(see Fig.~\ref{fig:cvi})} of \twCO\jtwo\ CVI over the
whole field. The statistics is significantly improved compared to
previous works.  The probability density in the most extreme bins
reaches \dix{-5}.  Fig.~\ref{fig:cvi} displays the locus of the E-CVI.
It is remarkable that the thin structure delineated by the
\twCO\jtwo\ E-CVI in the SE area (blue box of Fig.~\ref{fig:tdv}) is
so similar to that obtained with the same method applied to a much
smaller sample observed in a different transition,
\twCO\jone\ (Paper~II). The E-CVI structure is the high-angular
resolution view of the structure obtained with the same statistical
analysis performed on KOSMA maps of the field (HPBW=120\arcsec) and
shown in Fig.~12 of Paper~II.

The E-CVI structure does not follow the peaks of the \twCO\jtwo\ line
integrated emission. Instead, it coincides, over the $\sim 1$~pc
region discussed in Section 3.1, with the narrow interface of the HV
and LV components \ie\ the intense velocity-shear (Fig.~\ref{fig:cvi},
center), and follows in detail the thin elongated structure in the
extreme velocity range $[-2.0,-0.5]$~\kms, that of the red line\-wings
(Fig.~\ref{fig:cvi}, right).  This association between CO line\-wings
and intense velocity-shears extends the findings of Paper~II to
higher-resolution and over a larger scale.

These properties of the locus of E-CVI unambiguously support, for the
first time, the proposition of Paper~II that the E-CVI trace intense
velocity-shears in turbulent gas and that the extreme variations of
the line CV are driven by the appearance/disappearance of linewings
over small scales.  It is also the first observational proof of the
early conjecture of \cite{falgarone1990} that the broad CO linewings
trace the intermittency of turbulence in molecular clouds.  These
results clarify, at least in the case of translucent clouds, the
controversy on the origin of small-scale CV variations expected to be
due primarily to density fluctuations, line-of-sight projections, and
radiative transfer \citep{esquivel2007,miville2003b,levrier2004}.

Last, this E-CVI structure is coherent over $\sim2$~pc while its
thickness is as small as 0.03~pc. Its aspect ratio is therefore $\sim
70$. Its length seems to be limited by the size of the map (see the
longer structure computed from the KOSMA data in Paper~II).  We also
note that the E-CVI structure splits into multiple branches in several
areas, in particular around offsets $(-1000\arcsec, 800\arcsec)$ and
$(-700\arcsec, 500\arcsec)$.

\section{Discussion }

\subsection{What is the nature of the interface?}

The interface is primarily an intense velocity-shear.  The
\pv\ diagrams show that this velocity shear corresponds to a
discontinuity in the CO flow: the HV (LV) component is not detected
above $\sim0.5$~K in the SW (NE) of the shear.  The flows undetected
in the \twCO\jtwo\ line are either CO-poor and/or too dilute to excite
the transition.  In this framework, we observe the yield of a strain
developing in a gas undetected in the \twCO\jtwo\ line: the gas we
detect (denser and/or richer in CO) is generated in the 1-dimensional
intense-shear interface and is spread in the POS by motions whose
velocity cannot be measured.  This scenario naturally produces the two
components of the large velocity-shear, with sharp edges closely
following each other over $\sim1$~pc and little overlap in projection
{\it under any viewing angle}.

The intense-shear structure may however belong to a shock of unknown
velocity in the POS. We have searched for SiO\jtwo\ line emission as a
chemical shock signature within this structure and found no emission
above a significant low threshold $3 \sigma = 5$~mK that corresponds,
in the optically thin case, to a tiny SiO column density of about
\dix{10}~\cc.  Hence, there is no chemical signature of C-shocks
faster than 20~\kms\ detected at the scale of 0.03~pc
\citep{gusdorf2008}. But we cannot rule out a weak C-shock component
($v_S\leq 2$ \kms) in the POS that would produce the density
enhancement and/or the CO enrichment of the gas required to explain
the non-detection in the \twCO\jtwo\ line of the gas before it enters
the shear layer. This would be consistent with the sub- to
trans-Alfv\'enic nature of the turbulent motions in that field (see
Paper II).  The solenoidal contribution to the interface (2~\kms)
would exceed the possible compressive one ($\leq 2$ \kms), in
agreement with \cite{federrath2009} who find that our observed
statistical properties of turbulence in the Polaris Flare are in very
good agreement with solenoidal forcing at large scale.  \answer{This
  result is reminiscent of the finding of \cite{mininni2006a} that the
  stronger the shear at large-scale, the more intense the
  intermittency of velocity increments at small-scale}.

\subsection{A plausible link with the dense cores}

The above findings are similar to those inferred from PdBI
observations (FPH09) of the small (1'~by~2') field shown in
Fig.\ref{fig:cvi} (left).  Velocity-shears as intense as
500~\kmspc\ are detected there over distances of 6~mpc, at the edge of
CO structures of velocities differing by several \kms. The PdBI field
is close to two low-mass dense cores \citep{heithausen2002},
interestingly located at the tip of the E-CVI structure
(Fig.~\ref{fig:cvi}, left).

The viscous dissipation rate of turbulence being proportional to the
square of the rate-of-strain \citep{landau_fluid}, it is tempting to
interpret the large increase of the velocity-shear, from
30~\kmspc\ (Fig.~\ref{fig:tdv}) to 500~\kmspc\ in the PdBI field, as
due to the development of an instability in the large-scale shear.
The growth of the instability splits the shear into small-scale and
more intense shears, thus increasing the local dissipation rate of
turbulence by two-orders of magnitude.
Clustering of small-scale structures of high strain-rate magnitude
(and therefore large dissipation) into structures of inertial
extension have been found in numerical simulations of incompressible
HD \citep{moisy2004} and MHD turbulence \citep{mininni2006b}.  One
then may speculate that these bursts of dissipation eventually lead to
the formation of low mass dense cores largely devoid of turbulent
energy, after an evolution of the gas that remains to be understood.
    
\section{Conclusions}

We have detected a 1-dimensional structure of intense velocity-shear
($\sim$ 15 to 30 \kmspc), coherent over $\sim$ 1~pc with a thickness
of only 0.03 to 0.15~pc.  This remarkable structure follows the
distribution of extreme \twCO\jtwo\ line-wings and coincides partly
with the locus of E-CVIs in the field. These findings support the
previous claim we made that, in translucent molecular clouds, E-CVIs
are tracers of extreme velocity-shears in interstellar turbulence, as
do the broad CO linewings.

This shear structure is proposed to be the source of {\it layers} of
CO-rich dense gas in a CO-poor (and/or dilute) gas component
experiencing the strain and not seen in \twCO\jtwo.  The shear is
likely to be the site of enhanced turbulent dissipation. We cannot
rule out an undetected shock component in the POS.

These results, in conjunction with the PdBI results of FPH09, stress
the coupling of small and large scales in interstellar turbulence,
over a dynamic range never reached before, \ie\ from 6~mpc to more
than 1~pc.  They support a framework in which trans-Alfv\'enic (but
supersonic) turbulence dissipates primarily in intense-shear layers
connecting the large-scales to mpc scales (or below).

We speculate that turbulence dissipation has been proceeding for a
longer time at the southern tip of the E-CVI structure than in the
northern part, leading to the formation of the two dense cores.

\begin{acknowledgements}
  The authors thank M.~Heyer for his clarifying and perceptive report
  on the original version of this Letter.  We are grateful to C.~Thum
  and H.~Wiesemeyer from IRAM for their indefectible support to the
  HERA facility. The authors acknowledge the hospitality of the Kavli
  Institute for Theoretical Physics (Grant No. PHY05-51164).
\end{acknowledgements}

\bibliographystyle{aa}


\end{document}